\documentclass[twocolumn,showpacs,preprintnumbers,amsmath,amssymb]{revtex4}

\usepackage{graphicx}
\usepackage{dcolumn}
\usepackage{bm}
\usepackage{subfigure}

\begin{document}

\title{Hierarchical organization of brain functional network during visual task}
\author{Zhao Zhuo$^{1}$}
\author{Shi-Min Cai$^{1}$}
\email{csm1981@mail.ustc.edu.cn}
\author{Zhong-Qian Fu$^{1}$}
\email{zqfu@ustc.edu.cn}
\author{Jie Zhang$^{2}$}
\email{jzhang080@gmail.com}

\affiliation{$^{1}$Department of Electronic Science and
Technology, University of Science and Technology of China, Hefei
Anhui, 230026, PR China \\
$^{2}$Centre for Computational Systems Biology, Fudan
University, Shanghai 200433, PR China}

\date{\today}

\begin{abstract}
In this paper, the brain functional networks derived from
high-resolution synchronous EEG time series during visual
task are generated by calculating the phase
synchronization among the time series. The hierarchical
modular organizations of these networks are systematically
investigated by the fast Girvan-Newman algorithm. At the
same time, the spatially adjacent electrodes
(corresponding to EEG channels) are clustered into
functional groups based on anatomical parcellation of brain
cortex, and this clustering information are compared to
that of the functional network. The results show that the
modular architectures of brain functional network are in
coincidence with that from the anatomical structures
over different levels of hierarchy, which suggests
that population of neurons performing the same
function excite and inhibit in identical rhythms.
The structure-function relationship further reveals that
the correlations among EEG time series in
the same functional group are much stronger than
those in different ones and that the hierarchical
organization of brain functional network may be a
consequence of functional segmentation of brain cortex.
\end{abstract}


\maketitle

\section{Introduction}
Human brain, which is consisting of ten thousand million neurons and
even more synapses, is perhaps the most complex system ever known.
Benefiting from the development of brain anatomy since the
nineteenth century, we now know that the neuronal elements of the
brain constitute an extremely complex structural network, which
supports a wide variety of cognitive functions and neural activities
\cite{Cajal1995,Swanson2003}. Recently, it has gained great
interests of scientists to investigate functional connectivity of
brain based on complex network theory in which the brain can
naturally be abstracted as a functional network. A brain functional
network can be extracted based on functional MRI (fMRI),
electroencephalography (EEG), magnetoencephalography (MEG), or
multielectrode array (MEA) data, which record electric, magnetic, or
other signals representing cognitive cortical activities of brain
\cite{Bullmore2009}. Vertices of brain functional network derived
from fMRI data describe anatomically localized regions of interest
(ROIs) or voxels of fMRI image, whereas the vertices of those
derived from EEG, MEG, or MEA data mimic the surface electrodes or
sensors. The functional connectivity (or edge) between pairs of
vertices is usually estimated using correlation between records of
different vertices.

It has been widely observed that the brain functional networks
demonstrate properties such as small-worldness \cite{Watts1998} and
power-law degree distribution \cite{Barabasi1999}, which distinguish
brain functional network from regular and random networks
\cite{Stam2004,Eguiluz2005,Bassett2006,
Achard2006,Heuvel2008,Gong2009}. However, the small-worldness and
power-law degree distribution only represent the global properties
of brain functional networks. To understand brain functional
networks imposed by structural and functional constraints more
comprehensively, other aspects, which are expected to reflect both
local and global organization of brain functional network (e.g.
hierarchical modular organization), need to be investigated.

Hierarchical organization, also called community structure and
modular architecture, describes that some nodes in a network are
densely connected as groups, and these groups are only sparsely
connected among themselves, which is a common phenomenon in diverse
networks such as World Wide Web, scientist collaborating networks,
genetic networks, protein-protein interaction networks, and
financial networks
\cite{Newman2004,Wilkinso2004,Jonsson2006,Cai2010}. A large number
of algorithms are developed to detect the hierarchical organizations
of real networks \cite{Girvan2002,Newman2004,Capocci2005,Arenas2008,Zhang2009,Zhang2010}
(also see review work \cite{Fortunato2010}). Interestingly, several
recent works have also found the hierarchical organization of
brain functional network derived from resting-state fMRI data
\cite{Salvador2005,Ferrarini2009} and epileptic MEG signals
\cite{Chavez2010}, respectively.

Since the large-scale neuronal networks of brain emerge
from synchronized delta oscillations, the cortical EEG time
series are able to distinguish the ongoing from the evoked
activities of brain \cite{Langheim2006,Lu2007}. In our
work, the brain functional networks are derived from
high-resolution synchronous EEG time series, which
consist of 238 channels and are recorded during a cognitive
task involved in visual, judgment and motor functions of
brain. In particular, vertices correspond to surface
electrodes (i.e. channels), and edges are determined by
correlation (degree of phase synchronization) of EEG time
series from pairs of channels. In fact, the electrodes can
also be clustered into functional groups by their
spatial positions on the scalp and specifically priori
knowledge of anatomical parcellation of brain cortex. For
example, electrodes near visual cortex area mainly
represent visual function, and are therefore
clustered together. The anatomical parcellation of brain
cortex is performed according to Brodmann
segmentation scheme \cite{Brodmann}. The fast Girvan-Newman
(GN) algorithm is first applied to analyze the hierarchical
organizations of brain functional networks. Strong evidence
for the existence of modular architecture is found.
Comparing the clustering results from the brain functional
network and the anatomical segmentation, we find
significant coincidence of the modular architecture in the
network derived from EEG data and anatomical organization
of the cortex over different levels of hierarchy.
This result suggests that vertices are more
tightly coupled in same functional cortex region than those
belonging to different ones and that the
patterns of neural activities of brain cortex are to a
large extent determined by the anatomical modular
architectures of the brain.

\section{Materials and methods}
\subsection{Data acquisition}
The high-resolution EEG time series were synchronously
recorded during visual task by using a large number of
scalp electrodes (238-channels). Therefore, they had a high
spatial and temporal resolution, which provide very helpful
and detailed information of electrical activity of cortical
surface. Specifically, the single-subject data set was
recorded by A. Delorme et al. in the Swartz Center, UCSD,
with the sampling rate 256 Hz using a Biosemi Active Two
system \cite{web}. Experiments were performed as follows:
Filled white disks appeared briefly inside one of five
empty squares, and one of the five outlines was colored
green to mark the square as a visual target, then the
candidate made a motor response by pressing a mouse button
with their right hand as quickly as possible whenever the
filled white disk appeared at the attended location. It is
noted that these locations were counterbalanced across
blocks in pseudorandom order. The more detailed description
of experiment can be found in several previous
works \cite{Makeig2002,Makeig2004}. In this data set, each
visual target was represented by a synchronous EEG
recording with five sections, and about three
thousands trials of experiment were performed. In addition,
there are 235 channels of EEG used in our work by dropping
3 channels of EOG.

\subsection{Network construction based on Phase synchronization}

The concept of phase synchronization is introduced to study
synchronization of coupled oscillators and has gained particular
interests to investigate coupling among nonlinear complex systems
\cite{Kuramoto1984,Yeung1999,Timme2002}. The phase $\Phi(t)$ of a
real-value time series $X(t)$ is defined using Gabor's analytic
signal approach \cite{Gabor1946}:
\begin{equation}
V(t)=X(t)+iX_{h}(t)=Ae^{i\Phi(t)}, \label{equ:Hilbert}
\end{equation}
where the imaginary part $X_{h}(t)$ is the Hilbert transformation of
$X(t)$. Hence, the degree of phase synchronization between
time series $X_{i}(t)$ and $X_{j}(t)$ is evaluated by a
bivariate phase coupling index \cite{Schelter2006}:
\begin{equation}
R_{i,j}=\|\frac{1}{T}\sum_{t=1}^{T}e^{i(\phi_{i}(t)-\phi_{j}(t))}\|,
\label{equ:index}
\end{equation}
which is in the interval [0,1]. If the phases of two time
series are completely synchronized, the phase
coupling index will be maximum.

The phase synchronization method is able to reduce the
nonstationary effect, compared with the calculation
of the correlation of two EEG time series directly. The
functional connectivity (edge) between pairs of electrodes
(vertices) has been estimated using the measure of phase
synchronization, and then it is thresholded to
generate functional networks. Herein each vertex is
connected to its $N$ nearest neighbors
(i.e. those channels that are most
phase-synchronized with itself) \cite{xiaoke}. Notice that the resulting
brain functional networks are not necessarily symmetric,
i.e. if vertex $i$ is a neighbor of vertex $j$, vertex $j$
may not necessarily be a neighbor of vertex $i$,
and vice versa. Though the edges are intrinsically
directed, we analyze the networks in an undirected way for
convenience. The mean degree of network is
generally determined by $N$. However, since the
directed edges are identified as undirected ones,
the actual mean degree is a little larger than
$N$. In this way, the connectivity of brain functional
network is naturally guaranteed without dense connections,
which will otherwise disturb the hierarchy
detection.

\subsection{Clustering EEG channels by electrode position and anatomical parcellation of brain cortex}

EEG time series recorded through each channel mainly
represent the electrical activities of neurons near
the corresponding electrode. Thus, the EEG
channels can be directly clustered into functional
groups according to the spatial position of electrodes and
anatomical parcellation of brain cortex based on Brodmann
area. We used the Brodmann template image distributed with
MRIcro , which is restricted to the standard MNI space
\cite{MRIcro}. Resolution of image is
$181\times217\times181$ and size of voxels is
$1mm\times\times1mm\times1mm$. Each hemisphere is
partitioned into 41 areas according to cytoarchitecture of
neurons, together with the same labels suggesting same
cognitive functions in two hemispheres. The locations of
electrodes are registered to MNI space using SPM8 toolbox
(open-source software) \cite{SPM8}. By checking Brodmann
areas in which the electrodes belong to, the
electrodes that are spatially nearby are labeled by the
same Brodmann areas index. At last, these 235 channels are
clustered into 25 functional groups corresponding
to 25 Brodmann areas (e.g. functional group 47 corresponds
Brodmann area 47). Other Brodmann ares are absent due to
deeply inside position of the brain cortex.
The size of functional groups varies from 1 to 29,
as shown in Tab.\ref{tab:numberofchannels}. Moreover,
Brodmann areas can be roughly scaled into 9 major
substructures according to specific physiological
functions, by which the EEG channels can
be further clustered at a higher hierarchy, as shown in
Tab.\ref{tab:functionalgroup}. Hence, the EEG
channels are organized into two levels of hierarchy
according to the above mentioned segmentation.
\begin{table}
\caption{\label{tab:numberofchannels} Number of channels restricted to Brodmann area.}
\begin{ruledtabular}
\begin{tabular}{llll}
Brodmann &   EEG channel &  Brodmann &   EEG channel  \\
area&       number &      area&       number \\
\hline
20  &  29  &   4  &   7\\
    19  &  26  &  10 &    7\\
     6  &  19  &   8  &   5\\
    18  &  16 &   22  &   5\\
     9 &   14  &  38  &   5\\
    37  &  13 &   44  &   5\\
    45  &  12  &  43  &   4\\
    40  &  11  &  48  &   4\\
     7  &  10  &   1   &  3\\
    21  &   9  &   2   &  3\\
    46  &   9  &   5  &   2\\
    17   &  8  &  47   &  1\\
    39  &   8  &      &  \\
\end{tabular}
\end{ruledtabular}
\end{table}

\begin{table}
\caption{\label{tab:functionalgroup} Number of channels restricted to functional substructure}
\begin{ruledtabular}
\begin{tabular}{lll}
Functional &   Brodmann &   EEG channel  \\
substructure&        area&       number     \\
\hline
Broca's(B)    &44, 45        &17         \\
Audition(A)   &22           &5          \\
Cognition(C)  &9, 10, 46, 47   &31         \\
Emotion(E)    &38           &5          \\
Vision(V)     &17, 18, 19     &50         \\
Vision-parietal(Vp)   &7, 39         &18 \\
Vision-temporal(Vt)   &20, 21, 37     &51 \\
Motor(M)              &4, 6, 8        &31 \\
Sensory(S)            &1, 2, 5, 40     &19 \\
\end{tabular}
\end{ruledtabular}
\end{table}

The hierarchical organization of EEG channels have two
aspects, modular architecture of brain functional networks
detected by fast GN algorithm and functional groups
according to anatomical parcellation of brain cortex based
on Brodmann area. Thus, the association of these two kinds
of hierarchical organizations, e.g. the overlap between
the communities of the channel sets and the
functional groups, is a key question which is
expected to shed lights onto the relation between brain
function and structure.

\section{Empirical results}

\begin{figure}[!t]
\centering
\includegraphics[width=3.5in]{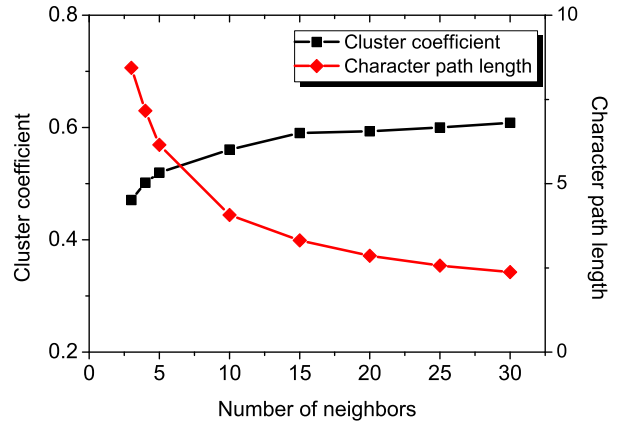}
\caption{(Color online) Average cluster coefficient and
character path length as a function of $N$ }\label{fig:smallworld}
\end{figure}

We calculate the phase-synchronization index matrix of all
the sections which consist of continuously recorded
multi-channel EEG time series. An average is performed over
all twenty-five sections. Then the brain functional
networks are generated using an alterable $N$ (the
number of nearest neighbors). The average cluster
coefficient (CC) and character path length
(CPL) are computed as a function of $N$ (see in Fig.
\ref{fig:smallworld}). The brain functional networks behave
large CC and short CPL, which suggests the small-world
property in consistency with previous works.

\begin{figure}[!t]
\centering
\includegraphics[width=3.5in]{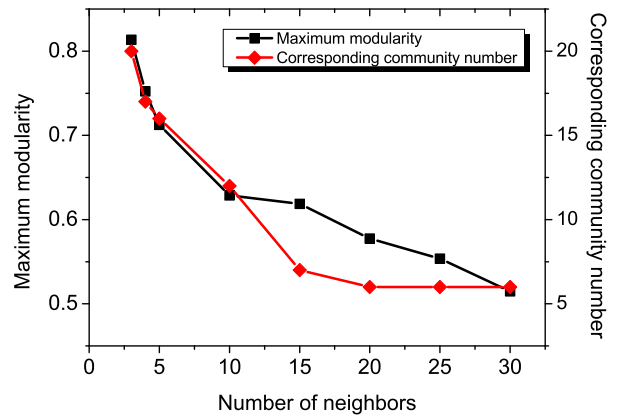}
\caption{(Color online) Maximum modularity and corresponding
community number as a function of $N$}\label{fig:comnum}
\end{figure}

Community structure and modular architecture are two
crucial properties of brain functional networks during
visual task. The fast GN algorithm is applied to
explore these hierarchical organizations of these networks.
Results are shown in Fig. \ref{fig:comnum} for different
$N$, of which the minimum value is $3$. Figure
\ref{fig:comnum} concretely shows that maximum modularities
monotonously decrease with $N$, all above 0.5 even
when $N \leq 30$, which obviously differs from that of
randomly connected network.
Furthermore, small maximum modularity also corresponds to
large mean degree of network, which suggests that the more
edges bridge the communities to reduce gaps of clusters and
render community structure less visible.
Obviously, the number of community corresponding to maximum
modularity is also a monotonously decreasing function of
$N$. With these considerations in mind, we mainly
investigate the community structure and modular
architecture of brain functional network with $N=3$.

\begin{figure}[!t]
\centering
\includegraphics[width=3.5in]{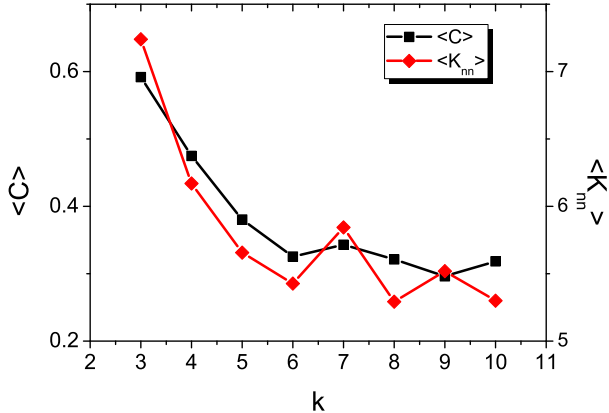}
\caption{(Color online) The distributions of $\langle C \rangle$ and
$\langle k_{nn} \rangle$ as a function of $k$.
Network is generated with $N=3$.}\label{fig:correlation}
\end{figure}

To better understand the architecture structure of brain
functional network, we study the average cluster
coefficient $\langle C \rangle$ for vertices with
degree $k$. Their relation reveals a negative correlation,
which suggests that low-degree vertices generally belong to
well connected clusters while the neighbors of high-degree
vertices belong to many different communities which are not
directly connected among themselves, namely hierarchical
organization (see Fig. \ref{fig:correlation}).
Simultaneously, the assortative mixing pattern
(degree-degree correlation of vertices) is investigated by
using a measure of average nearest-neighbor degree $\langle
K_{nn} \rangle$ that defined as the average over
vertices with degree $k$ (see Fig.
\ref{fig:correlation}). Degree mixing can be organized into
two patterns: assortative behavior if $\langle K_{nn}
\rangle$ increases with $k$, which indicates that
high-degree vertices are preferentially connected with
other high-degree vertices, and disassortative behavior if
$\langle K_{nn} \rangle$ decreases with $k$, which denotes
that links are more easily built between high-degree
vertices and small ones. In Fig. \ref{fig:correlation},
we find that $\langle K_{nn} \rangle$ decreases
with $k$, which indicates a disassortative
behavior of brain functional network.

\begin{figure}[!t]
\centering \subfigure[]
{\label{fig:modularityk3}\includegraphics[height=1.6in,width=3.5in]{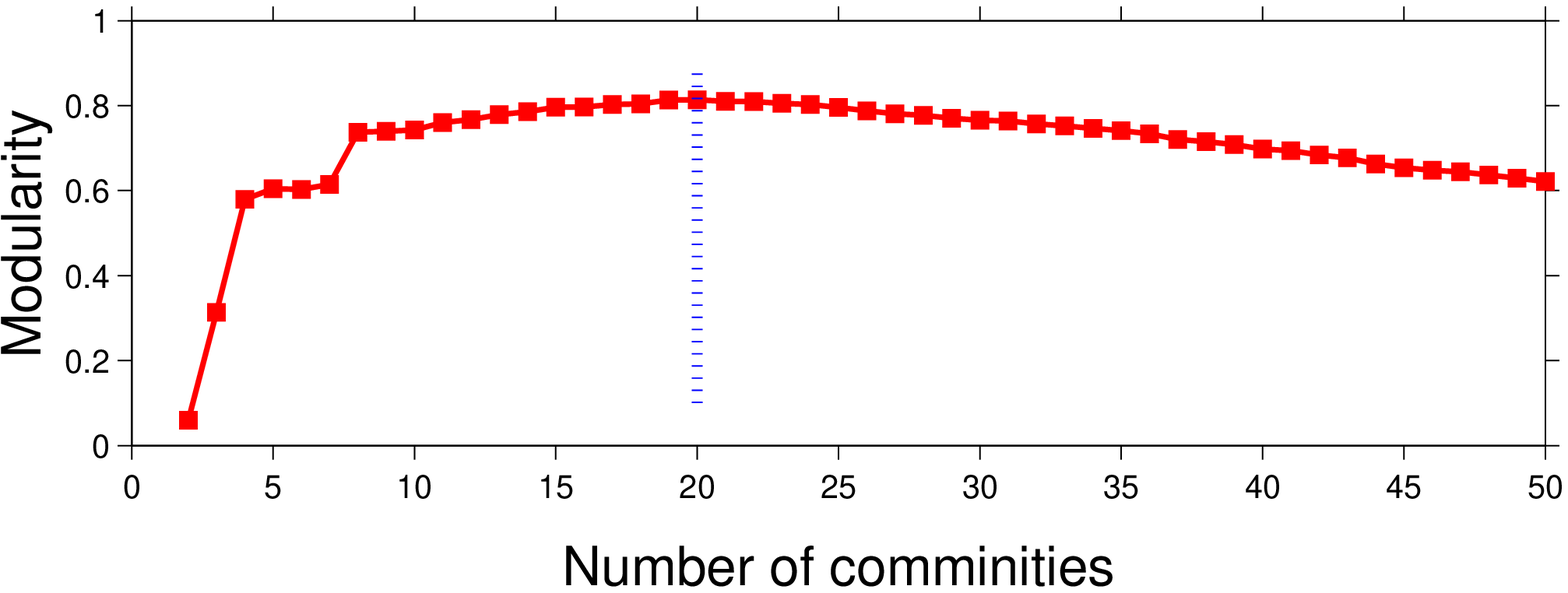}}
\hspace{0.3in} \subfigure[]
{\label{fig:Hierarchyk3}\includegraphics[height=1.6in,width=3.5in]{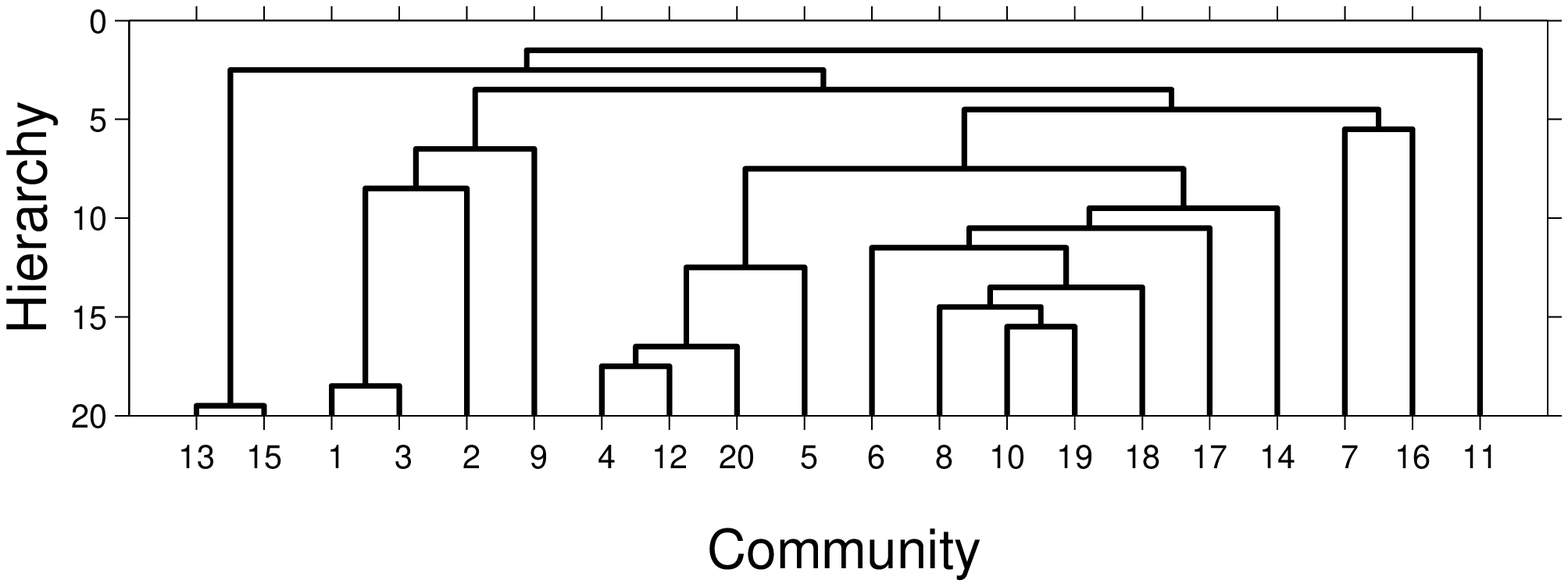}}
\caption{(Color online) Modularity (a) and dendrogram plot
(b) when network is generated with $N=3$} \label{fig:MM}
\end{figure}

The community structure of brain functional network is
further described by a dendrogram plot computed
with fast GN algorithms, as shown in Fig \ref{fig:MM}. In
Fig. \ref{fig:modularityk3}, the number of community
corresponding to maximum modularity is 20, which is marked
by a dash-line. Thus, in Fig. \ref{fig:Hierarchyk3}, the
dendrogram plot only shows hierarchical tree that splits
the network into 20 communities. Note that we do
not present the whole dendrogram plot so that the end
points of hierarchical tree still denote communities with
numeric labels from $1$ to $20$ that are randomly
ordered in Fig. \ref{fig:Hierarchyk3}.

\begin{figure}[!t]
\centering \subfigure[\ Composition of community
]{\label{fig:Composition}\includegraphics[width=3.5in]{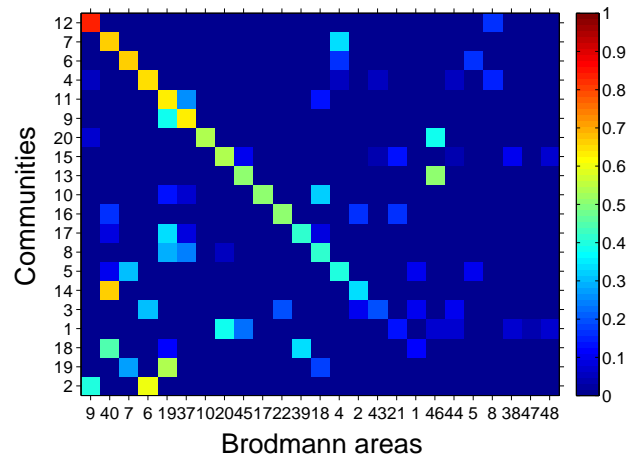}}\\
\subfigure[\ Participation of functional group]
{\label{fig:Participation}\includegraphics[width=3.5in]{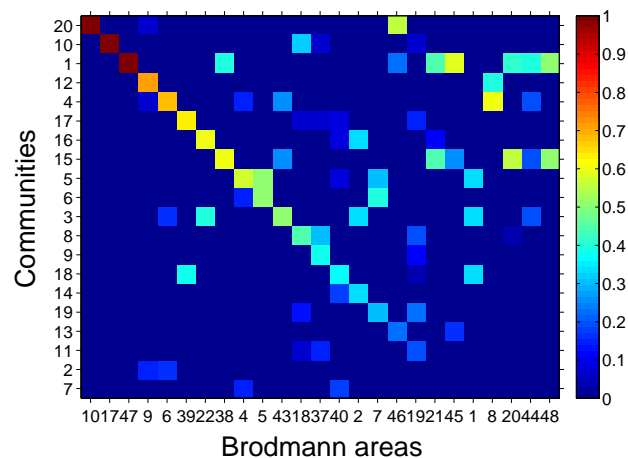}}
\caption{(Color online) (a) Composition of community from different
function groups restricted to Brodmann areas. (b) Participation of
functional group in communities of brain functional network. The
result reveals the coincidence between modular architectures of
brain functional network and functional groups based on anatomy
parcellation of brain cortex. Network generated with $N=3$ is
divided into 20 communities.} \label{fig:BrodmannCom}
\end{figure}

To further investigate the coincidence of the
two kinds of hierarchical organizations, we compute the
overlap between modular architecture of brain functional
network and functional groups based on anatomy parcellation
of brain cortex in two ways. On one hand,
we check how many EEG channels of same community
belong to the same function group, which is
represented by the composition of community written as
\begin{equation}
c(i,j) = |C(i) \cap G(j)|/|C(i)|. \label{equ:composition}
\end{equation}
On the other hand, we investigate whether EEG
channels of the same functional group are exactly
divided into an identical community, which is
described by participation of functional group defined as
\begin{equation}
p(i,j) = ||C(i) \cap G(j)|/G(j)|. \label{equ:participation}
\end{equation}
In Eqs. \ref{equ:composition} and \ref{equ:participation},
$C(i)$ and $G(j)$ denote the channel set of community $i$
and functional group $j$, respectively. If there is
a perfect one-to-one correspondence between communities and
functional groups, we will have $c(i,i)=p(i,i)=1$ and
$c(i,j)=p(i,j)=0$. Figure \ref{fig:BrodmannCom} shows the
composition of the communities, which suggests
that most of communities are mainly formed by channel sets
that are restricted to the same Brodmann area. For
instance, the community $12$ (numeric label) almost
overlaps with Brodmann area $9$, i.e. the channels
belonging to the same function group (by anatomical
parcellation) are densely connected, or coupled in the
corresponding brain functional network. In Fig.
\ref{fig:Participation}, the participation of functional
group reveals that most of channel sets restricted to
the same Brodmann areas are divided into
the same communities. In particular, Brodmann
areas $10$, $17$, and $47$ are completely overlapped with
the communities $20$, $10$, and $1$ (numeric labels),
respectively. It is noticed that some channel sets of
functional groups are divided into several large
communities because Brodmann areas involved with these
functional groups are distributed in both left and
right cerebral hemisphere with much larger spatial distance
which reduces the coupling strength among neurons.
For instance, Brodmann area 21 is mainly divided into two
larger communities with numeric labels $1$ and $15$, which
are distributed in left and right cerebral hemispheres,
respectively. This explains why the one-to-one
correspondence between communities and functional groups
are not perfect (i.e. $c(i,i),p(i,i)\neq 1$).

\begin{figure}[!t]
\centering \subfigure[\ Composition of community
]{\label{fig:Compositionh}\includegraphics[width=3.3in]{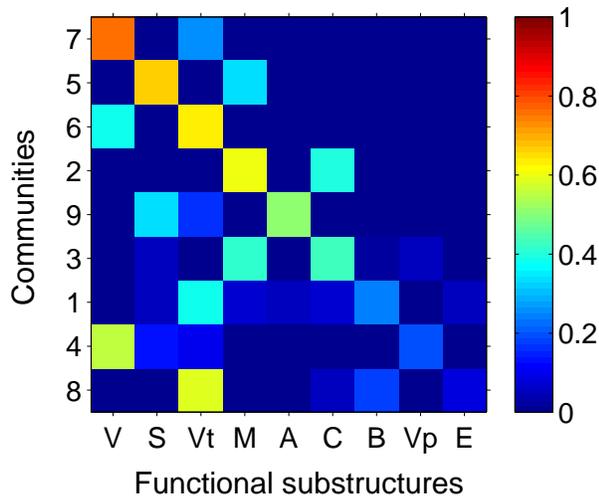}}\\
\subfigure[\ Participation of functional substructure]{\label{fig:Participationh}
\includegraphics[width=3.3in]{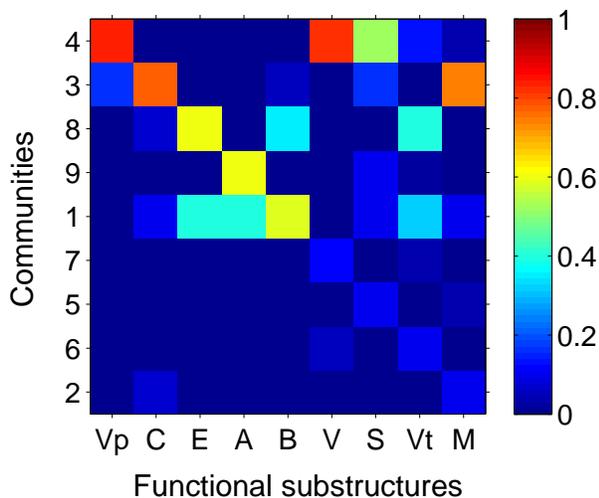}}
\caption{(Color online) (a) Composition of community from different
functional substructure restricted to specific physiological
functions. (b) Participation of functional substructure in
communities of brain functional network at a higher lever of
hierarchy. The result shows the coincidence between modular
architectures of brain functional network and functional
substructure according to specific physiological functions of brain
cortex. Network generated with $N=3$ is divided into 9 communities.}
\label{fig:GroupCom}
\end{figure}

In a higher level of hierarchy, the coincidence of modular
architectures (i.e. larger communities) of brain functional
network and functional substructures (see Tab.\ref{tab:functionalgroup}) 
of brain cortex is also studied
in aforementioned ways. Concretely, the brain functional
network is partitioned into 9 larger communities. The
composition of the communities shows that the
communities are mainly formed by channel sets
associated with functional substructures like
\emph{Vision}, \emph{Sensory}, \emph{Vision-temporal},
\emph{Motor}, which is well consistent with the visual task
involved in visual, judgment and motor functions of brain
(see Fig. \ref{fig:Compositionh}). For the
participation of functional substructure, most of
functional substructures are almost divided into unique
communities, except for the \emph{Visual-temporal}
part, which is symmetrically distributed in two hemispheres
and is separated into two communities, see Fig.
\ref{fig:Participationh}.

In addition, the electrical activity of neuronal networks
is known to oscillate at various frequencies and
amplitudes. The densely interconnected oscillators in same
communities will synchronize more easily than those
in different communities, which suggests that the
synchronization reveals hierarchical organization for a
network with a nontrivial community structure
\cite{Arenas2006,Zhou2006a,Gardenes2007,Arenas2007,Yan2007}.
Moreover, the simulation of neuronal activity based on
anatomical cat brain network determines that the correlated
clusters are consistent with anatomical areas of same brain
functions \cite{zhou2006b,Zemanova2006,zhou2007c}. Thus,
the coincidence of modular architectures of brain
functional network and function groups of brain cortex at
two levels of hierarchy suggests that the correlation of
EEG time series in same functional groups is much stronger
than that in different ones and the hierarchical
organization of brain functional network may be a
consequence of functional segmentation of brain cortex.

\section{Conclusion}
In conclusion, we have investigated the hierarchical
organization of brain functional network derived from
high-resolution synchronous EEG time series during visual
task through defining vertices as EEG channels and
evaluating connectivity between channels using a measure of
phase synchronization. The resulting brain functional
networks show common small-world property and community
structure organized in hierarchical way. Meanwhile,
by clustering EEG channels into functional groups based on
anatomical parcellation of brain cortex, we find that the
modular architectures of brain functional networks are in
coincidence with these functional groups at different
levels of hierarchy via computing overlap between
communities of channel sets and functional groups.
These interesting results suggest that population of neurons
performing the same functions excite and inhibit in
identical rhythms. This is reflected by that the
correlation between pairs of EEG time series (channels)
representing the neuroelectrical activity of same
functional cortex region is enhanced during tasks
related to vision, judgement and motor functions
of brain. The structure-function relationship further
reveals that the strong connection of channels and
community formation in brain functional networks may be a
consequence of functional segmentation of brain cortex.

\begin{acknowledgements}
This work is supported by the National Natural Science Foundation of China under
Grant Nos. 60874090, 60974079, 61004102. S-MC appreciates the financial support of
K.C. Wong Education Foundation, China Postdoctoral Science Foundation, and
the Fundamental Research Funds for the Central Universities.
\end{acknowledgements}

\end{document}